# Application of Grey System Theory in Approximate Calculation of Wave Packet Evolution[*]

Jiayu NI[1], Linfeng YE[1], Dehong GAO[1] and Xiangdong FENG [2][†]

[1] *Society of Theoretical and Computational Physics, College of Engineering and Technology of CDUT, Leshan 614000, China*
[2] *Basic Teaching Department, College of Engineering and Technology of CDUT, Leshan 614000, China*

The study of wave packet is very important in quantum mechanics, optics and fluid mechanics. However, in order to solve the rigorous evolution behavior of wave packet, it is necessary not only to determine the parameters of various physical quantities, such as mass, but also to carry out the complicated integration process of configuration space and momentum space. The evolution behavior of wave packet is the evolution behavior of parameters of distribution function with time in the final analysis. By using the method of grey system theory, the time-dependent parameters can be replaced by time response series, and then the formal structure can be made according to the physical meaning of time evolution, so as to realize the approximate simulation of wave function expression and carry out approximate calculation. The advantage of this method is that it can simulate the evolution of wave packet under the premise of uncertainty of each physical quantity. Based on the grey system theory, this paper replaces the time response function of GM (1,1) model for wave packet distribution, obtains the simulated evolution behavior of wave packet's wave function through mathematical techniques such as variational method, and makes error analysis and correction. In practical application, the approximate evolution behavior of wave function can be obtained only by determining the parameters of its development coefficient. In this paper, Gauss type wave packet is taken as an example to discuss the method. The grey system theory in mathematical modeling is introduced into the calculation of physics for the first time, and an approximate calculation method is provided, which can be widely used in the calculation of wave packet evolution.

## 1. INTRODUCTION

Wave packet is essentially a special solution of wave equation, because most of it is concentrated in a certain region, so that wave packet can show certain properties of particles. Because the object of wave mechanics is many partial differential equations, it has great advantages in numerical calculation and simulation. Therefore, on the one hand, for the study of particles, we tend to regard them as the superposition of a series of waves, that is, wave packets. It is a widely used physical model in the fields of quantum mechanics, optics and fluid mechanics [1-3]. On the other hand, in the wave behavior of the field, those solutions with obvious particle properties are related to the hot topic in recent years -- soliton problem [4-5]. The soliton is not an actual particle, but a wave with remarkable particle property. For example, scattering behavior and energy exchange can occur between solitons.

---

[*] Project supported by the Miao Zi project fund of College of Engineering and Technology of CDUT (C232020035)
[†] Corresponding author. E-mail: 759564613@qq.com

However, the strict expression of wave packet is often very difficult to solve, and some involve complex complex function integral. Even if the exact expression can be written, it is still not conducive to the subsequent calculation. In this case, the study of wave packet often needs to take numerical analysis method [6-8]. Grey system theory, as a theory to study its internal relationship by using data change trend, has become a powerful tool [9-13]. In this paper, Gauss wave packet in quantum mechanics is taken as an example to illustrate how the grey system theory is applied to approximate calculation of wave packet evolution.

The Gauss wave packet in quantum mechanics has the following wave function expression in the initial state

$$\psi(x,0) = (2\pi)^{-\frac{1}{4}} \sigma^{-\frac{1}{2}} e^{-\frac{(x-x_0)^2}{4\sigma^2}} \tag{1}$$

It satisfies Gauss distribution. Generally speaking, it is the superposition of several monochromatic plane waves, which can be expressed by Fourier transform pairs in position $x$ space and momentum $p$ space. The time evolution behavior can be expressed by adding a time evolution factor (in this paper, the physical quantity under the integral sign represents the integral of the quantity on $(-\infty,\infty)$)

$$\begin{aligned}\psi(x,t) &= \frac{1}{\sqrt{2\pi\hbar}} \int_p \psi(p) e^{\frac{i}{\hbar}p\cdot x - \frac{i}{\hbar}\frac{p^2}{2m}t} dp \\ &= \frac{1}{2\pi\hbar} \int_x \int_p \psi(0,0) e^{\frac{i}{\hbar}p\cdot x - \frac{i}{\hbar}\frac{p^2}{2m}t} dp\, dx \end{aligned} \tag{2}$$

In this way, the wave function expression of Gauss wave packet can be obtained.

Using generalized Fresnel integral

$$\int_{-\infty}^{\infty} dx \cdot e^{iax^2} = \sqrt{\frac{i\pi}{a}} \tag{3}$$

we can calculate the integral of the momentum space, and then calculate the integral of the position space, so as to obtain the complete wave function of Gauss type wave packet

$$\psi(x,t) = \sqrt[4]{\frac{1}{2\pi}} \sqrt{1/\sigma[1+i\hbar t/2m\sigma^2]} \exp\left\{\frac{im(x-x_0)^2}{2(\hbar t - i2m\sigma^2)}\right\} \tag{4}$$

We can let

$$\sigma(t) \equiv \sigma\sqrt{1 + \frac{\hbar^2 t^2}{4m^2\sigma^4}} \tag{5}$$

to further simplify eq.(4). For convenience, let $\kappa = \hbar/2m\sigma^2$, and

$$\psi(x,t) = \sqrt[4]{\frac{1}{2\pi}}\,\sigma(t)^{-\frac{1}{2}} \exp\left\{-\frac{(x-x_0)^2}{4\sigma(t)^2}(1 - \mathrm{i}\kappa t) - \frac{\mathrm{i}}{2}\arctan(\kappa t)\right\} \tag{6}$$

Eq.(6) describes the free evolution of a one-dimensional Gauss type wave packet with mean square error $\sigma(t)$, and its initial mean square error is $\sigma(0) = \sigma$.

Although the evolution law of Gauss type wave packet can be solved according to the theory, this work is very complicated in practice, which requires not only knowing the mass $m$ of the particle, but also performing more complex integration. Is it possible to replace the time-dependent parameters $\sigma$ of the initial Gauss type wave packet eq.(1) with the time response function in the grey system theory, and then construct the imaginary index part of the time evolution, so as to approximate the evolution law of the wave packet?

We know that the free evolution of Gauss type wave packet is that the wave packet gradually becomes "fat", that is, $\sigma(t)$ is monotonically increasing, and the change of wave packet is monotonic. This creates conditions for us to model the grey system of eq.(6) and solve the time response series, so as to approximate the evolution law of Gauss type wave packet with time.

## 2. INTEGRATION OF GREY SYSTEM THEORY

### 2.1 Time Response Substitution of Mean Squre Error

It seems that the final state wave packet (6) is formed by replacing the mean square deviation $\sigma \to \sigma(t)$ of the initial state wave packet (1) and multiplying it by the imaginary index part representing the time evolution. For the replacement of $\sigma$, there are many mature models in grey system theory, and the most widely used one here is GM(1,1) model [14-16]. The time response function of GM(1,1) is $\sigma_G(t) = (\sigma - C)\mathrm{e}^{-\alpha t} + C$. Where $\alpha$ is called the development coefficient and $C$ is the undetermined constant. We divide the exponential part of the wave function of equation (6) into two parts

$$\begin{aligned}\psi(x,t) &= \sqrt[4]{\frac{1}{2\pi}}\,\sigma(t)^{-\frac{1}{2}} \exp\left\{-\frac{(x-x_0)^2}{4\sigma(t)^2}\right\} \\ &\quad \exp\left\{-\frac{\mathrm{i}}{\hbar}\left(\frac{\hbar}{2t}\arctan(\kappa t) - \frac{(x-x_0)^2}{4\sigma(t)^2}\kappa\hbar t\right)t\right\}\end{aligned} \tag{7}$$

The reason why we are divided into two lines is that it is convenient to compare with the form of stationary wave function $\psi(x,t) = \psi(x)\mathrm{e}^{-\mathrm{i}Et/\hbar}$, so it is convenient for us to construct the imaginary exponential function. We hope that we can fully simulate the distribution of the whole wave function on the $(x,t)$ only by taking the development coefficient $\alpha$ as a parameter.

The reason why we are divided into two lines is that it is convenient to compare with the form of stationary wave function $\psi(x,t)=\psi(x)\mathrm{e}^{-\mathrm{i}Et/\hbar}$, so it is convenient for us to construct the imaginary exponential function. We hope that we can fully simulate the distribution of the whole wave function on the $(x,t)$ only by taking the development coefficient $\alpha$ as a parameter.

For the part of coefficient and real exponential function (the first line in eq.(7)), $\sigma(t)$ can be directly replaced by the time response function in grey system theory, which is equivalent to the function approaching problem

$$\sigma(t)=\sigma\sqrt{1+\kappa^2 t^2}\sim(\sigma-C)\mathrm{e}^{-\alpha t}+C=\sigma_G(t) \tag{8}$$

Just make them equal $\sigma\sqrt{1+\kappa^2 t^2}=(\sigma-C)\mathrm{e}^{-\alpha t}+C$ the development coefficient can be given, the second order small quantity is denoted as $\mathcal{O}[2]$

$$\alpha=\frac{\ln(\frac{\sigma-C}{\sigma\sqrt{1-\kappa^2 t^2}-C})}{t}=\frac{\kappa^2\sigma t}{2(C-\sigma)}+\mathcal{O}[2] \tag{9}$$

If we take the development coefficient to the right of eq.(8), we can get $\sigma_G(t)$. Next, we use the variational method to get the appropriate $C$ [17]. Consider the following functional

$$I=\int_0^\infty \mathrm{d}t\left(\sigma\sqrt{1+\kappa^2 t^2}-(\sigma-C)\exp\left\{-\frac{\kappa^2\sigma t^2}{2(C-\sigma)}\right\}-C\right) \tag{10}$$

Its physical meaning is quite obvious, that is, the integral of the error between $\sigma(t)$ and $\sigma_G(t)$ in the evolution time. We hope that this integral can be minimized, that is, the variation of the functional is zero. And because the parameter $\kappa,t$ is in the form of square product, it has symmetry in eq.(10), so only one of them needs to be processed in the variational operation. We choose to variational $t$ and make it zero. Finally we get

$$\frac{\kappa^2 t\sigma}{\sqrt{1+\kappa^2 t^2}}-\exp\left\{-\frac{\kappa^2 t^2\sigma}{2(C-\sigma)}\right\}\kappa^2 t\sigma=0$$
$$C=\sigma+\frac{\kappa^2 t^2\sigma}{\ln(1+\kappa^2 t^2)} \tag{11}$$

We take the result into eq.(9) and get the development coefficient $\alpha=\ln(1+\kappa^2 t^2)/2t$, constant $C=\sigma[1+(\mathrm{e}^{2\alpha t}-1)/2\alpha t]$, then the $\sigma_G(t)$ given by grey system theory is expressed as

$$\begin{aligned}\sigma_G(t)&=-\sigma\cdot\frac{\mathrm{e}^{2\alpha t}-1}{2\alpha t}\mathrm{e}^{-\alpha t}+\sigma(1+\frac{\mathrm{e}^{2\alpha t}-1}{2\alpha t})\\&=\sigma\left(1+\frac{\mathrm{e}^{2\alpha t}-1}{2\alpha t}(1-\mathrm{e}^{-\alpha t})\right)\end{aligned} \tag{12}$$

The approximate function of the first line can be obtained by replacing $\sigma(t)\rightarrow\sigma_G(t)$ in eq.(7).

**2.2 Constructive Approximation of Imaginary Index Part**

For the imaginary exponent part (the second line in eq.(7), we can use the physical meaning of the time evolution operator $e^{-iEt/\hbar}$, and make a special structure. Becausse of $\kappa = \hbar/2m\sigma^2$, $\kappa$ can characterize the wave packet quality $m$. If this wave is an electromagnetic wave in a vacuum and propagates at the speed of light, then it should satisfy Einstein's mass energy relation $E = mc^2$, substitute it into the time evolution operator, we will have

$$e^{-\frac{i}{\hbar}Et} = e^{-\frac{i}{\hbar}mc^2 t} = e^{-\frac{i}{\hbar}\frac{1}{\kappa}\frac{\hbar c^2}{2\sigma^2}t} \tag{13}$$

For other physical fields, the energy relation can be adjusted to $E = 1/2mv^2 + U$, where $m$ is the mass of the wave packet, $v$ is the propagation velocity of the wave packet, and $U$ is the potential energy given to the wave packet by the field. Then we construct an expression similar to eq.(13) to characterize the time evolution behavior of wave packet.

According to $\alpha = \ln(1 + \kappa^2 t^2)/2t$, we can write $E = \hbar t c^2/2\sigma^2\sqrt{e^{2\alpha t} - 1}$ constructed by $\alpha$. In this way, the imaginary exponential factor characterizing the time evolution of wave packet becomes the key factor

$$\exp\left\{-\frac{i}{\hbar}Et\right\} = \exp\left\{-\frac{ic^2 t^2}{2\sigma^2\sqrt{e^{2\alpha t} - 1}}\right\} \tag{14}$$

So far, we completely write out the distribution of Gauss wave packet approximation wave function on the surface determined by the development coefficient $\alpha$ and based on the grey system theory

$$\psi_G(x,t) = \sqrt[4]{\frac{1}{2\pi}}\sigma_G(t)^{-\frac{1}{2}}\exp\left\{-\frac{(x-x_0)^2}{4\sigma_G(t)^2} - \frac{ic^2 t^2}{2\sigma^2\sqrt{e^{2\alpha t} - 1}}\right\}$$

$$\sigma_G(t) = \sigma + \frac{e^{2\alpha t} - 1}{2\alpha t}(1 + e^{-\alpha t}) \tag{15}$$

For convenience, we call $\psi_G$ the gray wave function. And the coefficient and the real index are formally denoted as $\phi(x,t)$, then there is

$$\psi(x,t) = \phi \cdot \exp\left\{-\frac{i}{\hbar}\left(\frac{\hbar}{2t}\arctan(\kappa t) - \frac{(x-x_0)^2}{4\sigma(t)^2}\kappa\hbar t\right)t\right\}$$

$$\psi_G(x,t) = \phi_G \cdot \exp\left\{-\frac{ic^2 t^2}{2\sigma^2\sqrt{e^{2\alpha t} - 1}}\right\} \tag{16}$$

And in the following discussion, we need to note that $\kappa$ is only related to the type of experimental particles, which is essentially a constant. $x_0$ as the symmetry axis of Gaussian wave packet, we can take $x_0 = 0$ as the origin. Next, our work is to analyze the errors of $\psi$ and $\psi_G$ and make corrections.

## 3. ERROR ANALYSIS

In the above discussion, the original wave function eq.(7) and the approximate gray wave function eq.(15) in the gray system are given. In contrast, for the part of function $\phi$, $\phi$ is consistent with $\phi_G$, as long as the error between $\sigma_G(t)$ and $\sigma(t)$ is sufficiently small, then the error between $\phi$ and $\phi_G$ is also sufficiently small. To achieve this, we first consider the relative error of variance at time $t=0$

$$\lim_{t \to 0} \left| \frac{\sigma_G(t) - \sigma(t)}{\sigma(t)} \right| = 0 \tag{17}$$

The result shows that the relative error of $\sigma_G$ and $\sigma$ is zero at the initial time. With the evolution of time, we can judge the nature of the relative error function by analyzing the characteristics of the whole function. We define the relative error function

$$\sigma_{eff} = \left| \frac{\sigma_G(t) - \sigma(t)}{\sigma(t)} \right| \tag{18}$$

We can draw a rough image of $\sigma_{eff}$ as

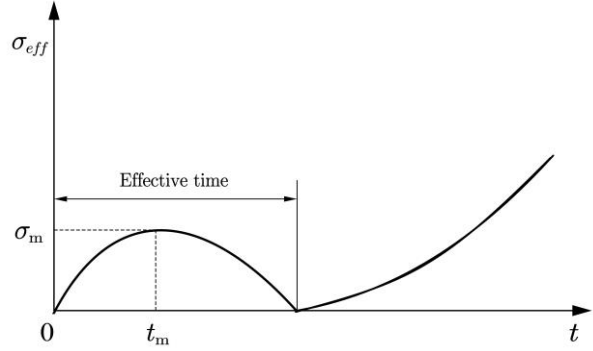

**Fig 1**. Schematic diagram of relative error function of variance. Defines the time from the initial time to the rightmost zero as the effective time, which is the main simulation time interval of grey wave function due to its controllable maximum value. According to its function characteristics, there is a maximum relative error in the effective time $\sigma_m$.

It can be seen that the image of relative error function rises all the way after reaching the rightmost zero point. From the initial time to this time, the relative errors of $\sigma_G$ and $\sigma$ are within a controllable range. We take this period of time as our simulation time interval, which is called effective time. In the effective time, the relative error of variance has a maximum point $t_m$. It is not easy to solve the algebraic expression $t_m = t_m(\kappa, \alpha)$ of $t_m$. This is mainly because we are going to use the method of local extremum to solve it. But when $\alpha$ is very small, there is more than one arch structure in the effective time, so there is more than one local extremum. But we can get $t_m$ by solving the zero point of relative error function $\sigma_{eff}$, limiting the range of time variable to the rightmost arch structure, and then solving the local maximum at this time, so as to determine the location of the maximum error. This method will be demonstrated in the example analysis in Section 4.

Let's analyze that $t_m$ is obviously related to the parameters $\kappa, \alpha$, and $\kappa$ only depends on the type of particles, so it can be said that $\alpha$ is determining the maximum error. If we require the error, for example, the simulation error should not be higher than 10%, then we can get an appropriate $\alpha$ by solving the programming problem $\sigma_{eff}(t_\mathrm{m}) \leqslant 0.1$ to bring into $\phi(x,t)$. At the same time, the effective time as the zero value of the maximum relative error function $\sigma_{eff}$ will also be determined. It can be proved that the lower the maximum error in the effective time is, the shorter the effective time is. According to our requirements, we will require $\sigma_{eff}(t_\mathrm{m})$ to meet the error requirements, and the effective time is as long as possible.

## 4. APPLICATION EXAMPLE AND DISTANCE EFFECTIVENESS

### 4.1 An Example of Electromagnetic Wave

We take a long wave commonly used in electromagnetic technology as an example. The photon dynamic mass of this long wave is about $3.2 \times 10^{-38}$ kg, and the wavelength is about $1000$ m [18]. It can be obtained by numerical calculation that when $\kappa = 0.035$, the wave function at $t=0$ time can better meet the experimental measurement results of the electromagnetic wave. Considering $\kappa = \hbar/2m\sigma^2$, this means that the initial variance $\sigma \approx 216.5$. Then take this as an example to test our grey system method.

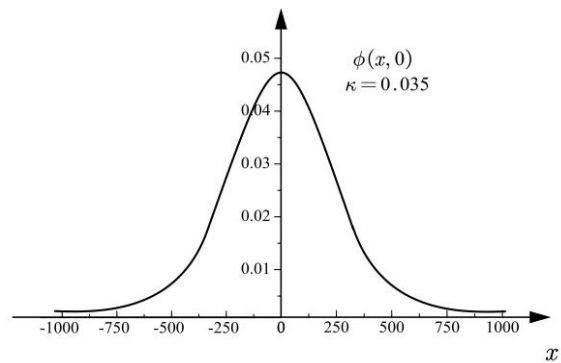

**Fig 2**. The wave function image of a photon. After calculation, the part falling in the interval $[-1000, 1000]$ accounts for 99.8% of the whole.

According to the error analysis technique proposed above, we now need to solve the appropriate development coefficient $\alpha$ according to the known $\kappa$ and the requirement that the error should not exceed 10%, and make the effective time as long as possible. So the problem comes down to a planning problem

$$\begin{cases} \sigma_{eff}(t_m) \leqslant 0.1 \\ \dfrac{\partial \sigma_{eff}}{\partial t}\bigg|_{t_m} = 0 \\ \sigma_{eff} = \dfrac{1}{\sqrt{1+\kappa^2 t^2}} + \dfrac{(1-e^{-\alpha t})(e^{2\alpha t}-1)}{2\alpha t \sqrt{1+\kappa^2 t^2}} - 1 \\ \kappa = 0.035 \end{cases} \quad (19)$$

In this way, according to eq.(19), we can use appropriate methods (such as Monte Carlo method [19-20]) to get the maximum error of 9.9% and the effective time of 95.43 s when the maximum error time $t_m = 60.5$ s and the development coefficient $\alpha = 0.01165$, which is the best result to meet the conditions. At this time, we take $\alpha$ into $\phi_G(x,t)$ and $\phi(x,t)$ to calculate the relative error. In time, taking 5 s as the unit, we draw the relative error between the gray wave function and the wave function at $x = 0$ m, 300 m, 500 m, 1000 m as the comparison

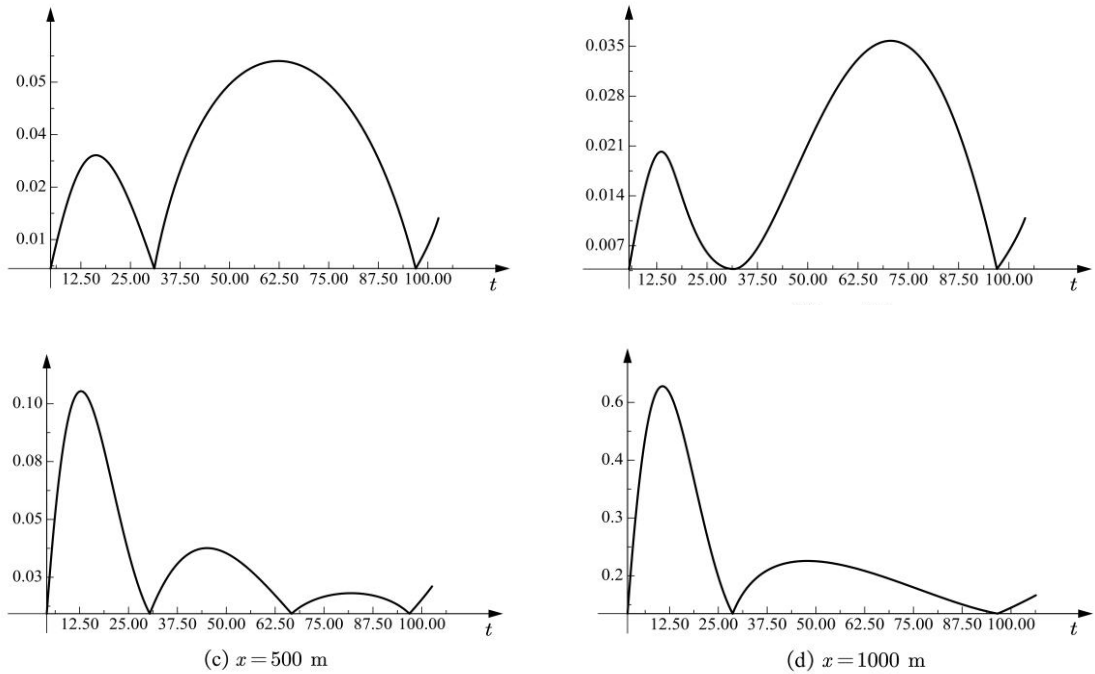

Fig 3. Gray wave function and relative error image of wave function at 0, 300, 500, 1000 m. It can be seen that the error of the whole effective time is not big in the distance of (a) and (b). But in the distance of (c) (d), the error exceeds 10% of the maximum value.

The (a) and (b) in Fig.3 tell us that the model is completely applicable within a certain range, the relative error is very small, and the effective time is long enough. But (c) (d) tells us that in a long distance, the model begins to lose its effectiveness and the error is too large. This shows that in addition to the validity of time, the model we built also has the validity of distance, that is, the concept of effective distance, which will be explained in detail in Section 4.2.

We can see that the result of programming problem eq.(19) contains rich mathematical structure. Now, we want to discuss the relationship between the effective time and the error. Since the maximum error needs to solve more complex equations, we take the average relative error $y$ in the whole effective time as the evaluation error

$$y = \frac{\int_0^{t_{eff}} \sigma_{eff}(t)\mathrm{d}t}{t_{eff}} \tag{20}$$

Now using the numerical analysis technology [21], the $\alpha - t_{eff}$ diagram and $\alpha - y$ diagram are drawn as follows

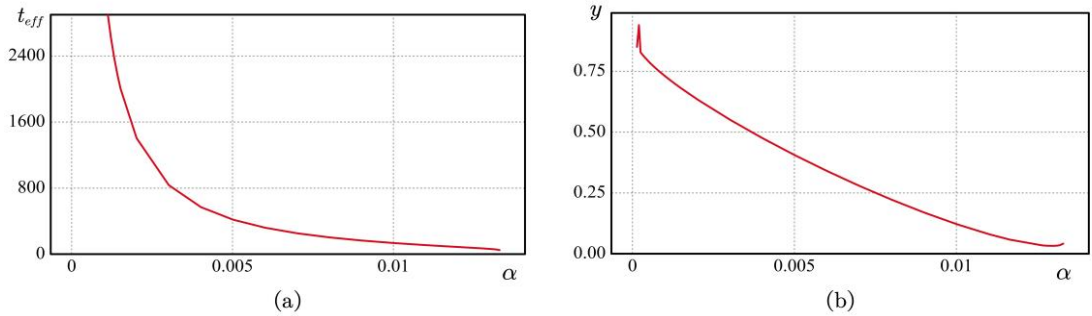

**Fig 4**. $\alpha - t_{eff}$ diagram and $\alpha - y$ diagram. The (a) shows that the effective time decreases rapidly with the increase of development coefficient. And (b) shows that with the increase of the development coefficient $\alpha$, the average relative error in the effective time decreases continuously.

Fig.4 shows that the development coefficient g does have a suitable value to meet our requirements for simulation error. When $\alpha$ is too large, although the average relative error in the effective time is small, the effective time is also small. On the contrary, when $\alpha$ is small, although the effective time is large, the average relative error in the effective time is also large. Therefore, selecting an appropriate $\alpha$ according to the simulation requirements is the fundamental guarantee for the effectiveness of the model. Because of this, when selecting $\alpha$, we can use function fitting and other methods, and then combine with algebraic means to solve the appropriate $\alpha$, which will not be repeated here [22].

**4.2 Application Examples and Distance Effectiveness**

So far, we have successfully established an effective model based on grey system theory to simulate the wave solutions of any physical field. But for the wave function itself, there are still some parts except $\phi(x,t)$, which can be used to further limit the validity of the model established above. The following is a detailed description.

For the rest, reference eq.(16) has

$$\varphi = \psi/\phi = \exp\left\{-\frac{i}{\hbar}\left(\frac{\hbar}{2t}\arctan(\kappa t) - \frac{(x-x_0)^2}{4\sigma(t)^2}\kappa\hbar t\right)t\right\}$$

$$\varphi_G = \psi_G/\phi_G = \exp\left\{-\frac{ic^2 t^2}{2\sigma^2\sqrt{e^{2\alpha t}-1}}\right\}$$

(21)

Where $\sigma(t)$ is defined by eq.(5). Noted that $\varphi_G(t)$ is only related to $t$, but $\varphi(x,t)$ is distance dependent, which indicates that the gray wave function $\psi_G(x,t)$ has an effective limit on the distance of the whole wave function $\psi(x,t)$ simulation. We need to understand that the error between $\varphi_G$ and $\varphi$ is not important. When considering various physical meanings, we only need to consider the simulation of $\phi_G$ to $\phi$ as in Section 4.1. Therefore, the significance of comparing $\varphi_G$ and $\varphi$ is only to provide a distance limitation on the effectiveness of simulation.

We first plot the relative difference $\Delta = (\varphi_G - \varphi)/\varphi$ between $\varphi_G$ and $\varphi$ at the end of the effective time

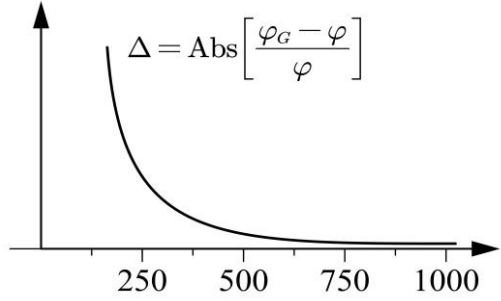

**Fig 5**. The schematic diagram of the relative difference between $\varphi_G$ and $\varphi$ at the final time $t = 95.43$ s of the effective time.

From the preliminary conclusion of the effectiveness of the model discussed above, the farther the distance, the lower the effectiveness of the model, and the greater the relative error between the gray wave function and the wave function. However, the farther $\Delta(x,t)$ is, the smaller its change rate is, which shows that the smaller $\Delta(x,t)$ is, the less effective the model is, and we can use the change rate of $\Delta(x,t)$ to get the effective distance. Consider the extremum of the second derivative

$$\frac{d^2}{dx^2}\Delta(x, 95.43) = 0$$

(21)

a solution $x = 484.5$ m satisfying the condition is obtained by numerical method, which is the effective distance of the model. Simulation beyond this distance will exceed the required error as shown in Fig.3 (c). This is the result obtained by using the part of the wave function $\varphi$. we take this result into the part of $\phi$ to verify it. It is found that the maximum relative error between the gray wave function and the wave function at $x = 484.5$ m is very close to 10%. When $x > 484.5$ m, the maximum relative error will exceed 10%, which destroys the effectiveness of the model to a certain extent. Finally, we find that the model established above is applicable to

$0 \text{ m} \leqslant x \leqslant 484.5 \text{ m}, 0 \text{ s} \leqslant t \leqslant 95.43 \text{ s}$ in the specific example of Section 4.1.

## 5. SUMMARY AND DISCUSS

In physics research, there are a lot of scenarios that depend on numerical calculation. However, grey system theory, a powerful tool in the field of mathematical modeling, has not been widely used in these calculations. In this paper, the grey system theory is applied to the simulation of wave function, a specific method is proposed, a complete mathematical model covering time and space is established, and its error properties are analyzed in depth.

However, due to the limited energy and level, the method of establishing mathematical model proposed in this paper can still be improved and discussed in depth. For example, in Section 4.1, we find that when the average relative error decreases, the effective time also decreases rapidly. In practical application, the requirement of error may lead to too small effective time and unsatisfactory model. In this case, we can use a mathematical trick, adding parameter in the replacement $\lambda: \sigma \to \lambda \cdot \sigma_G$ to modify the relative error function, so that the effective time of simulation is longer under the condition of the same level error. Similar revision and expansion will be carried out in the follow-up work.

The work done in this paper, the introduction of mathematical modeling technology and ideas in physics research has a pioneering significance, and its feasibility has been proved in this paper. I sincerely hope that in some areas of physics where numerical calculation and approximation are needed, it can be organically combined with mathematical modeling to create greater work.